\documentclass{optica-article}
\usepackage{physics}
\journal{opticajournal} 

\articletype{Research Article}

\usepackage{lineno}

\begin{document}

\title{Enhanced orbital-angular-momentum dimensionality and spatial--spectral decoupling in thin-film SPDC}

\author{Dmitrii Akatev,\authormark{1,*} Jakub Lewandowski,\authormark{1,2} and Maria Chekhova\authormark{1,3}}

\address{\authormark{1}Max Planck Institute for the Science of Light, Staudtstr. 2, 91058 Erlangen, Germany\\
\authormark{2}Institute of Experimental Physics, University of Warsaw, 02-093 Pasterura 5, Warsaw, Poland\\
\authormark{3}Friedrich-Alexander Universität Erlangen-Nürnberg,  91058 Erlangen, Germany}

\email{\authormark{*}dmitrii.akatev@mpl.mpg.de} 

\begin{abstract*} 
Thin-film spontaneous parametric down-conversion, due to the relaxed phase-matching condition, generates photon pairs that are extremely broadband in frequency and angle and highly entangled in time/frequency and position/momentum. Here we show that it also provides a very broad orbital angular momentum (OAM) spectrum, which, in contrast to the case of bulk crystals, is decoupled from the frequency spectrum.  To demonstrate it experimentally, we generate photon pairs in a $7~\mu\mathrm{m}$-thick lithium niobate film and compare them with those from a $500~\mu\mathrm{m}$-thick periodically poled lithium niobate crystal under identical pump conditions. The thin-film source exhibits a substantially broader OAM spectrum, with the azimuthal Schmidt number almost 9 times larger than that of the bulk source. On this highly entangled state, we perform quantum-state tomography in selected two- and four-dimensional OAM subspaces, demonstrating high-purity maximally entangled states. These results highlight thin nonlinear films as a promising platform for high-dimensional quantum-state generation with weak spatial--spectral coupling.
\end{abstract*}

\section{Introduction}

Thin-film nonlinear films have recently emerged as versatile platforms for the generation of entangled photons through spontaneous parametric down-conversion (SPDC)~\cite{Review_flat_optic}. In films with thickness comparable to or smaller than the nonlinear coherence length of SPDC, the phase matching is satisfied automatically~\cite{MicroScaleGen, Stich2026,Sultanov22}. As a result, unlike bulk crystals, thin films can generate photon pairs under nonzero wavevector mismatch. This enables SPDC with large spectral and angular bandwidth and provides substantially greater flexibility in the choice of nonlinear material and operating wavelength. At the same time, the micro- or nanometer-scale thickness of such nonlinear media facilitates their direct integration with planar photonic platforms and heterostructures, providing a route toward compact on-chip quantum-light sources~\cite{Joshi2026,Choi2026}.

The giant angular bandwidth of thin-film SPDC provides access to a large number of spatial modes, forming high-dimensional photonic Hilbert spaces. This makes thin-film SPDC particularly attractive for spatial-mode encoding, including orbital-angular-momentum (OAM) entanglement and high-dimensional quantum information processing~\cite{Nape2023QuantumStructuredLight,Babazadeh2017HighDimensionalGates}. In a $d$-dimensional space, a single photon can encode up to $\log_2 d$ bits of classical information using $d$ orthogonal states, exceeding the one-bit capacity of a two-dimensional encoding~\cite{Holevo1973,Karimi_enhc_cap}.

Simultaneous access to spatial and spectral modes provides an additional resource for quantum-state engineering. In particular, OAM and frequency can be exploited jointly for multidimensional encoding, multiplexing, and the generation of
hyperentangled states~\cite{GarciaEscartin2008, Fracasso2020}. For these applications, however, the spatial and spectral degrees of freedom should ideally remain independently addressable. In conventional SPDC, this is not the case~~\cite{Baghdasaryan2022, Osorio2008}: the spatial and spectral degrees of freedom are generally coupled because both depend on the longitudinal phase matching. In thin films, the longitudinal phase-matching constraint is strongly relaxed, which leads to factorable, albeit extremely broad, frequency and angular spectra.

In this work, we experimentally demonstrate the substantially increased spatial dimensionality enabled by thin-film nonlinear sources. We directly compare the spiral (OAM) bandwidths of photon pairs generated in a $7~\mu\mathrm{m}$-thick lithium niobate film and a $500~\mu\mathrm{m}$-thick phase-matched lithium niobate crystal under identical pump conditions. We show that the thin-film source features a considerably broader OAM spectrum. Further, using the thin-film source, we perform quantum-state tomography of the generated two-photon states in three two-dimensional OAM subspaces and one four-dimensional subspace, revealing high-purity states with high fidelity to the corresponding maximally entangled target states. We further investigate spatial--spectral coupling of the generated OAM states. Numerical simulations show that the thin-film source preserves a nearly unchanged spiral spectrum over a broad spectral range, whereas the bulk crystal exhibits a pronounced frequency-dependent variation of the OAM distribution. Frequency-resolved OAM measurements provide the first experimental indication of spatial--spectral decoupling in the thin-film source. Together, these results demonstrate the potential of thin-film nonlinear crystals for broadband, high-dimensional quantum-state generation with reduced coupling between the spatial and spectral degrees of freedom, opening opportunities for spatial--spectral hyperentanglement.

\section{The OAM spectrum of a thin SPDC source}

In SPDC, a pump photon is annihilated and a pair of lower-energy photons is generated, conventionally called  signal and idler. With their transverse wavevectors denoted as $\mathbf{q}_s$ and $\mathbf{q}_i$,  the two-photon state is
\begin{equation}
\ket{\Psi}=\iint d
\mathbf{q}_s d\mathbf{q}_i \Phi(\mathbf{q}_s,\mathbf{q}_i)
\hat{a}^{\dagger}_s(\mathbf{q}_s)
\hat{a}^{\dagger}_i(\mathbf{q}_i)
\ket{0},
\label{eq}
\end{equation}
where $\Phi(\mathbf{q}_s,\mathbf{q}_i)$ is the spatial two-photon amplitude (TPA), determined by the transverse wavevector spectrum of the pump $E_p(\mathbf{q})$, the crystal thickness $L$, and the longitudinal wavevector mismatch $\Delta k_z = k_p - k_s-k_i$:
\begin{equation}
\Phi(\mathbf{q}_s,\mathbf{q}_i)
\propto
E_p(\mathbf{q}_s+\mathbf{q}_i)
\operatorname{sinc}\left(\frac{L\Delta k_z}{2}\right)
\exp\left(i\frac{L\Delta k_z}{2}\right).
\label{bip_amp}
\end{equation}
For a weakly focused pump, the angular spectrum $E_p(\mathbf{q}_s+\mathbf{q}_i)$ imposes strong transverse-wavevector anticorrelations, $\mathbf{q}_s+\mathbf{q}_i\approx0$.

For bulk nonlinear crystals, efficient SPDC generally requires the longitudinal phase-matching condition, $\Delta k_z=0$, to be satisfied. 
In contrast, for a nonlinear film with a thickness considerably smaller than the coherence length $L_{\mathrm{coh}}\equiv\pi/\Delta k_z$, the second factor in Eq.~(\ref{bip_amp}), known as the phase-matching function, is close to unity over a broad range of wavelengths, and  Eq.~\eqref{bip_amp} reduces approximately to
\begin{equation}
    \Phi(\mathbf{q}_s,\mathbf{q}_i)
    \propto
    E_p(\mathbf{q}_s+\mathbf{q}_i).
    \label{relax_phase_mathcing}
\end{equation}
Thus, for photon pairs generated in a thin film, the angular correlations are governed by the transverse profile of the pump, while the angular spectrum is broad and independent of the frequency spectrum due to the nearly uniform longitudinal phase-matching function~\cite{Sultanov22, MicroScaleGen}. 

The spatial structure of the state can be conveniently described in terms of Laguerre--Gaussian $\mathrm{LG}_{p}^{\ell}$ modes, where $p$ is the radial index and $\ell$ the azimuthal index associated with the OAM of the mode.  
Taking into account the OAM conservation, $\ell_p=\ell_s+\ell_i$, with the subscripts $p,s,i$ denoting the pump, signal, and idler photons, respectively, the spatial two-photon state can be written as
\cite{Miatto2012,Miatto2011}
\begin{equation}
    \ket{\Psi} = \sum_{\ell,p_s,p_i}C_{p_s,p_i}^{\ell}\ket{\mathrm{LG}_{p_s}^{\ell}}_s\ket{\mathrm{LG}_{p_i}^{\ell_p-\ell}}_i,
    \label{SPDC_full_LG_state}
\end{equation}
where $C_{p_s,p_i}^{\ell}, \sum_{p_s,p_i,\ell} |C_{p_s,p_i}^{\ell}|^2=1$, are mode amplitudes.

Since in the following we are interested only in the azimuthal structure of the generated photon pairs, the radial degrees of freedom are traced out. The corresponding marginal OAM probability distribution,
\begin{equation}
    P_\ell = \sum_{p_s,p_i}\left|C_{p_s,p_i}^{\ell}\right|^2,
    \qquad
    \sum_\ell P_\ell=1,
    \label{eq:OAM_probability}
\end{equation}
defines the OAM spectrum of the generated state. Its effective width can be quantified by the azimuthal Schmidt number \cite{Miatto2012Schmidt}
\begin{equation}
    K_{\mathrm{az}}
    =
    \frac{1}{\displaystyle\sum_{\ell}P_{\ell}^{2}}.
    \label{eq:Schmidt_1}
\end{equation}
In the thin-crystal approximation, where $L/(k_p w_p^2)\ll1$,
\begin{equation}
    K_{\mathrm{az}}\approx w_p\sqrt{\frac{k_p}{L}},
    \label{eq:Schmidt_scaling}
\end{equation}
where $w_p$ is the pump-beam waist and $k_p$ is the pump wave vector inside the nonlinear medium. Thus, the accessible azimuthal
dimensionality increases with the pump-beam waist and decreases with the square root of the crystal length, illustrating why thin nonlinear media can support substantially broader spiral spectra.

\section{Spatial--spectral decoupling of thin-film SPDC}

In the general case, the TPA in Eq.~\eqref{bip_amp} depends on both the transverse momenta and frequencies of the generated photons \cite{Baghdasaryan2022, SevillaGutierrez2024}. As a result, the OAM-mode distribution can vary across the SPDC spectrum, making the OAM spectrum frequency dependent. For a monochromatic pump, introducing the frequency detuning $\Omega$ from degeneracy, the full spatial--spectral TPA can be written as
\begin{equation}
\Phi(\mathbf q_s,\mathbf q_i,\Omega) \propto
E_p(\mathbf q_s+\mathbf q_i) \operatorname{sinc}\left[ \frac{L}{2}\Delta k_z(\mathbf q_s,\mathbf q_i,\Omega) \right]
\exp\left[ i\frac{L}{2}\Delta k_z(\mathbf q_s,\mathbf q_i,\Omega) \right],
\label{spec_spat_coupling}
\end{equation}
where
\begin{align}
\Delta k_z =k_{p,z}(\mathbf q_s+\mathbf q_i,\omega_p) - k_{s,z}\left( \mathbf q_s,\frac{\omega_p}{2}+\Omega \right) - 
k_{i,z}\left( \mathbf q_i,\frac{\omega_p}{2}-\Omega \right),
\end{align}
with
\begin{equation}
k_{j,z}(\mathbf q_j,\omega_j)
=
\sqrt{k_j^2(\omega_j)-|\mathbf q_j|^2},\,j=s,i, \omega_{s,i}=\omega_p/2\pm\Omega.
\end{equation}
For a Gaussian pump carrying zero OAM, 
$\ell_s+\ell_i=0$. The azimuthal structure of the photon pair is therefore determined by the dependence of the TPA on the relative azimuthal angle $\varphi=\phi_s-\phi_i$. The contribution $F_\ell(q_s,q_i;\Omega)$ of the OAM pair $(\ell,-\ell)$ to the two-photon amplitude at frequency detuning $\Omega$ is obtained by projecting the TPA onto the corresponding angular harmonic \cite{Miatto2012Spatial}:
\begin{equation}
F_\ell(q_s,q_i;\Omega)
=
\frac{1}{2\pi}
\int_0^{2\pi}
\Phi(q_s,q_i,\varphi,\Omega)
e^{-i\ell\varphi}\,d\varphi .
\end{equation}

Since we are interested only in the azimuthal degree of freedom, the radial degrees of freedom are traced out. The resulting frequency-resolved OAM weight is
\begin{equation}
\lambda_\ell(\Omega) = \int q_s\,dq_s \int q_i\,dq_i\, \left|F_\ell(q_s,q_i;\Omega)\right|^2.
\label{eq:lambda_l_Omega}
\end{equation}
The corresponding normalized frequency-resolved OAM spectrum is then 
\begin{equation}
P_\ell(\Omega) = \frac{\lambda_\ell(\Omega)} {\displaystyle\sum_m \lambda_m(\Omega)}.
\label{eq:P_l_Omega}
\end{equation}

In thin-film SPDC, 
the phase-matching function is nearly uniform over the relevant spectral and transverse-momentum ranges, strongly suppressing the intrinsic spatial--spectral coupling. The two-photon amplitude can be factorized as
\begin{equation}
    \Phi(\mathbf{q}_s,\mathbf{q}_i,\Omega)
    =
    \Phi_{\mathrm{sp}}(\mathbf{q}_s,\mathbf{q}_i)\,\Phi_{\mathrm{spec}}(\Omega).
    \label{eq:spatio_spectral_separable}
\end{equation}
The normalized OAM spectrum is then frequency-independent, $P_\ell(\Omega)= P_\ell$.
At the same time, signal and idler photons can remain entangled within both the azimuthal and frequency degrees of freedom. This provides a route toward spatial--spectral hyperentanglement, where the photon pair is entangled in both degrees of freedom while the spatial and spectral structures remain separable from each other. 

In the next section, we experimentally verify the broad azimuthal spectrum of thin-film SPDC and demonstrate its decoupling from the frequency spectrum.

\section{Experimental setup}

\begin{figure}[h]
    \centering
    \includegraphics[width=1\linewidth]{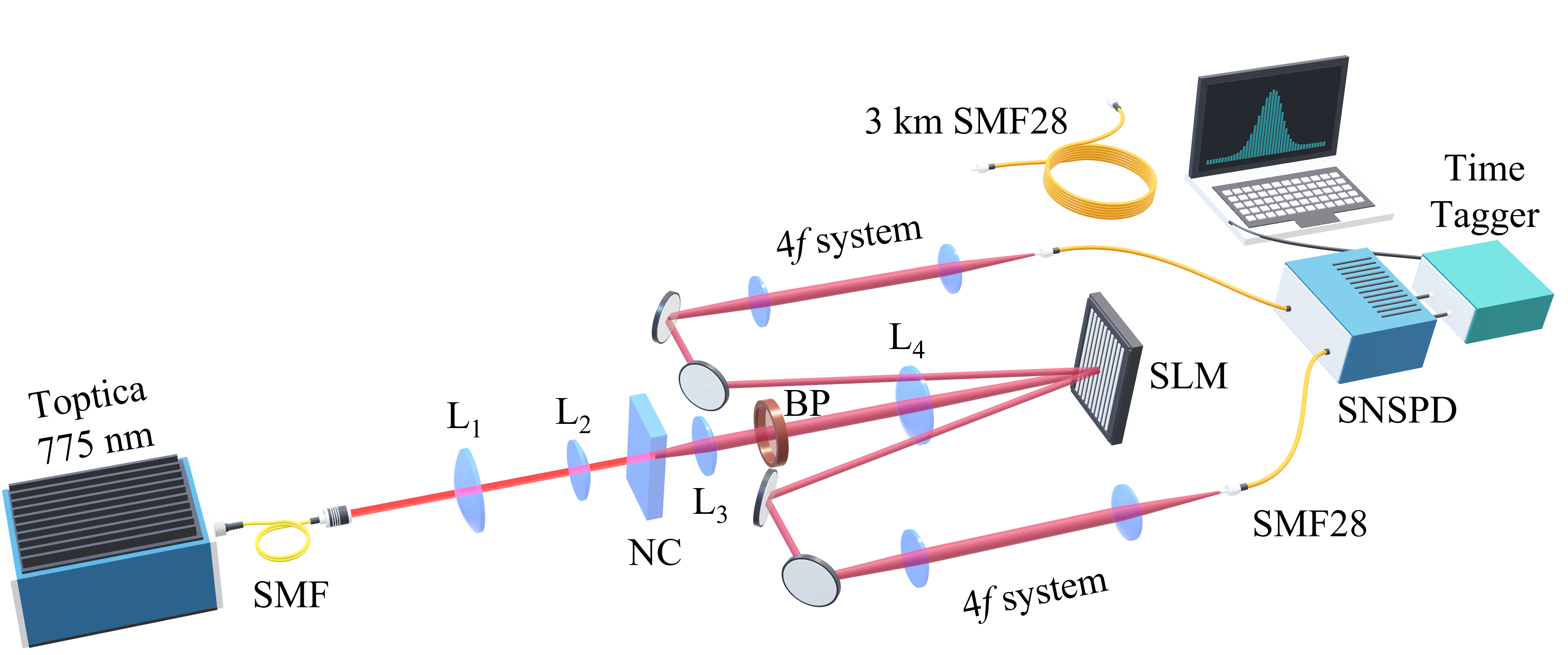}
    \caption{Schematic representation of the experimental setup used for SPDC measurements.} 
    \label{fig:setup}
\end{figure}

As a pump, we used a continuous-wave tuneable laser (Toptica DL pro) operating at 775 nm (Figure~\ref{fig:setup}) with 40 mW power. After mode-cleaning with a single-mode fibre, the beam was
 collimated and focused into a waist of $w_p = 37.5~\mu\mathrm{m}$ at the nonlinear crystal (NC) through a telescope formed by lenses $\mathrm{L}_1$ and $\mathrm{L}_2$.

Our thin-film sample was a $7~\mu\mathrm{m}$-thick, X-cut, MgO-doped lithium niobate (LN), where phase matching was satisfied automatically for a broad range of pump wavelengths $\lambda_p>680~nm$. For comparison, we used a $500~\mu\mathrm{m}$-thick periodically poled lithium niobate (ppLN) crystal with a poling period of $\Lambda=19.4~\mu\mathrm{m}$. In both samples, we generated photon pairs through type-0  SPDC. 

After spectral filtering with a 50~nm-wide band-pass filter centered at 1550~nm (BP), the photon pairs were directed to a spatial light modulator (SLM; HOLOEYE PLUTO-1, TELCO). A second telescope, implemented as a 4$f$ imaging system consisting of lenses $\mathrm{L}_3$ and $\mathrm{L}_4$, imaged the crystal plane onto the SLM. Because the fundamental-mode size in SPDC depends on both the crystal length and the pump beam waist \cite{Miatto2012}, the magnification of 4$f$ imaging system was optimized separately for each crystal. This ensured that, after the subsequent mode projection and imaging optics, the fundamental SPDC mode was well matched to the fundamental mode of the single-mode fiber used for photon detection, thereby maximizing the fiber-coupling efficiency and accessing to the full generated OAM spectrum.
Lens $\mathrm{L}_4$ had a focal length of $f=500~\mathrm{mm}$, whereas the focal length of $\mathrm{L}_3$ was selected separately for each nonlinear sample: $f=4.5~\mathrm{mm}$ for LN  and $f=11~\mathrm{mm}$ for ppLN.

The OAM spectrum of the generated photons was measured using the phase-flattening technique~\cite{PhaseFlatOAMEntanglement}. Signal and idler photons were spatially separated using a mode-multiplexing approach~\cite{Li2015OAMDemultiplexing}, in which the corresponding holograms directed them into the $+1$ and $-1$ diffraction orders, respectively.
After spatial separation into the signal and idler arms, the selected diffraction orders were coupled into single-mode fibers optimized for operation at 1550~nm (SMF-28). Additional 4$f$ imaging systems in the two arms, each providing a demagnification factor of approximately 140, imaged the corresponding modes onto the fiber-coupling planes.
The photons were detected using superconducting nanowire single-photon detectors (SNSPDs; Single Quantum). Photodetection pulses were recorded using a time-tagger (Swabian Instruments Time Tagger 20), from which  single-photon and coincidence count rates were extracted. An additional 3~km-long SMF-28 fiber was used for time-of-flight spectroscopy \cite{Valencia2002}, enabling frequency-resolved OAM detection.

\section{Results}

\subsection{The OAM bandwidth}

\begin{figure}[h]
    \centering
    \includegraphics[width=1\linewidth]{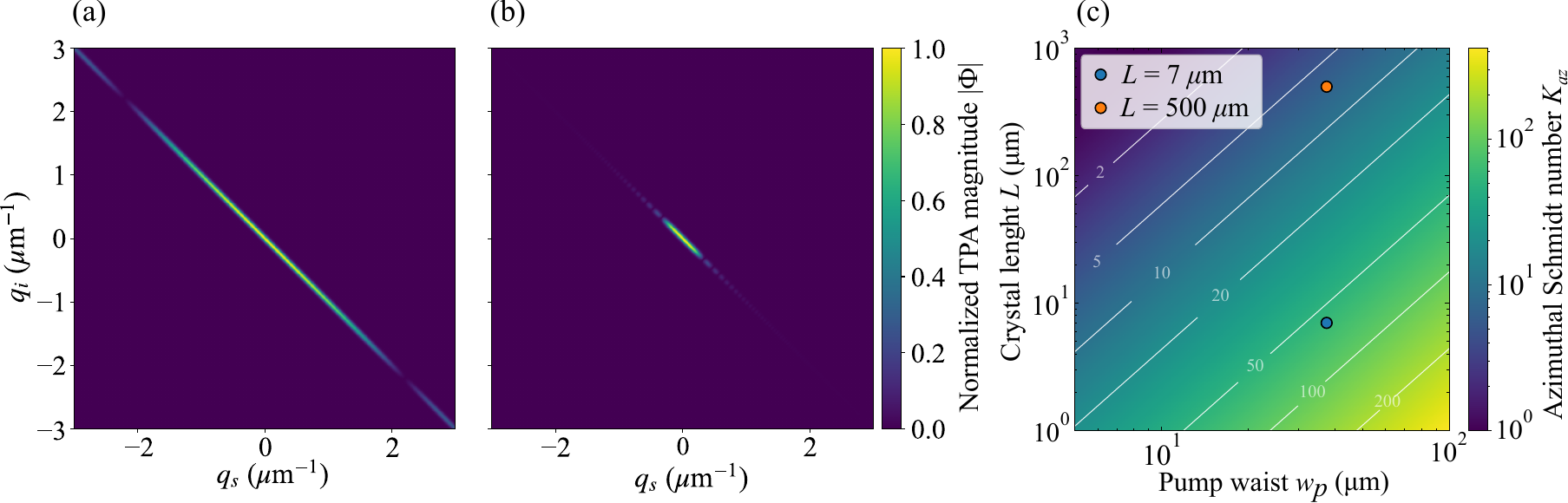}
    \caption{ Momentum-space two-photon amplitude, normalized to its maximum, for collinear degenerate SPDC pumped at $775~\mathrm{nm}$ in (a) a $7~\mu\mathrm{m}$-thick X-cut LN film and (b) a $500~\mu\mathrm{m}$-thick phase-matched ppLN crystal. (c) Azimuthal Schmidt number $K_{\mathrm{az}}$ as a function of the pump waist and the crystal length.}
    \label{fig:TPA_K_az}
\end{figure}

Figures~\ref{fig:TPA_K_az}(a,b) show the TPA in transverse-momentum space, calculated for both samples according to Eq.~\eqref{bip_amp}, with the pump-beam waist $w_p=37.5~\mu\mathrm{m}$, as in our experiment.  Figure~\ref{fig:TPA_K_az}(c) shows the dependence of the azimuthal Schmidt number on the crystal thickness and pump-beam waist. It is clearly seen that the thin-film LN supports a substantially broader transverse-momentum (angular) emission than the ppLN crystal, thereby providing access to a larger number of spatial modes that reflect in a bigger $K_{\mathrm{az}}$ value. Correspondingly, its spiral bandwidth is 8.9 times larger, with $K_{\mathrm{az}}^{\mathrm{LN}}=59.4$ compared to $K_{\mathrm{az}}^{\mathrm{ppLN}}=6.66$ under the same pump  (blue and orange circles in Figure~\ref{fig:TPA_K_az}(c), respectively).

\begin{figure}[h]
    \centering
    \includegraphics[width=0.9\linewidth]{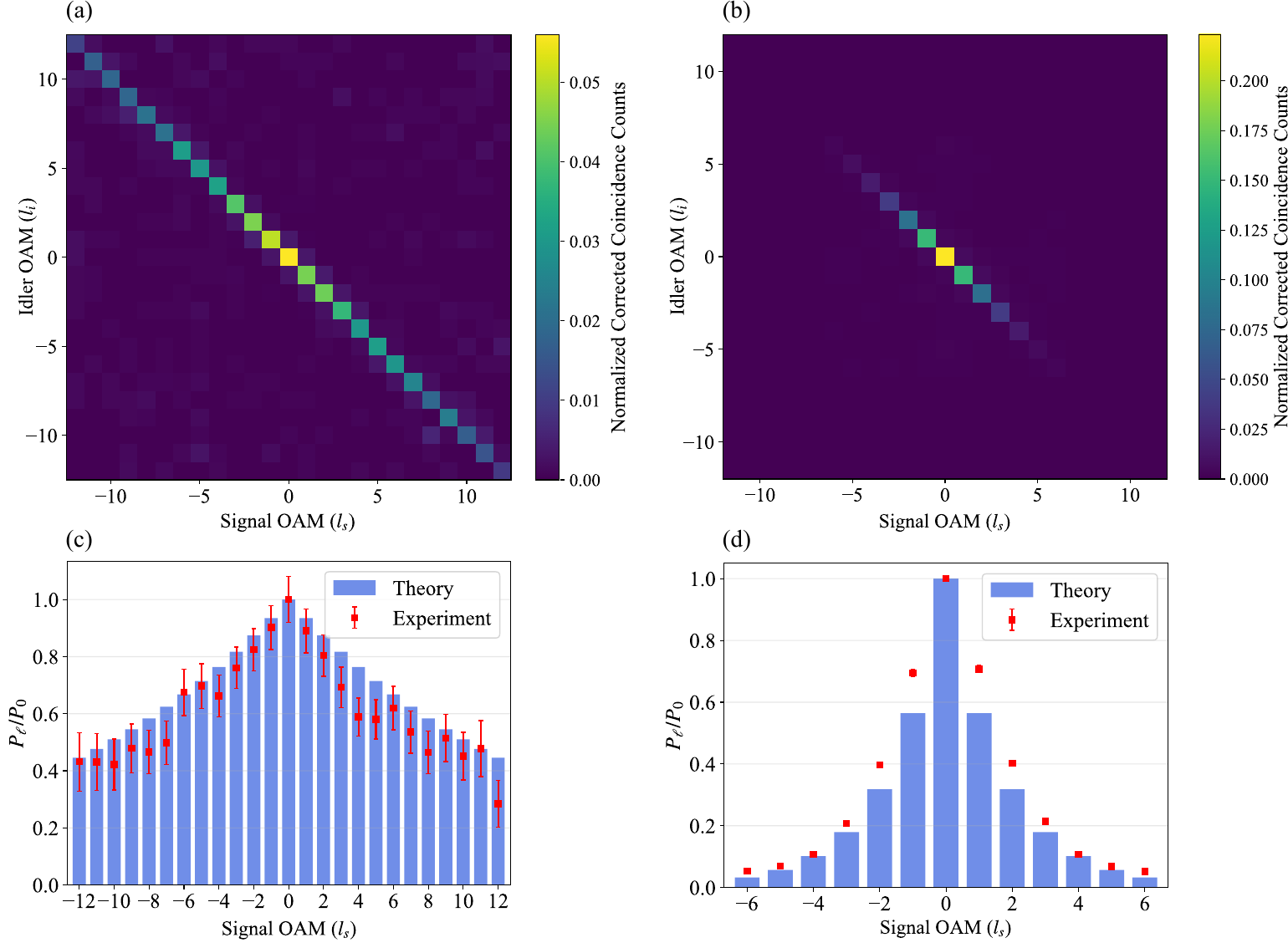}
    \caption{Measured joint OAM spectra of photon pairs generated by SPDC in (a) thin-film LN and (b) ppLN, together with the corresponding measured (red points) and calculated (blue bars) marginal distributions (c,d). The measurements were performed under optimal collection conditions at the nonlinear crystal.}
    \label{fig:OAM_spectrum_both}
\end{figure}

Figures~\ref{fig:OAM_spectrum_both}(a,b) show the measured joint OAM spectrum of photon pairs from the LN thin film (a) and from the ppLN crystal (b). Both distributions exhibit pronounced anticorrelation between the OAM charges of the signal and idler photons. The experimentally accessible
range was limited to $|\ell|\leq 12$ by the decrease in the SLM diffraction efficiency, $\eta_{\mathrm{diff}}$, at higher OAM orders.
Nevertheless, the thin film clearly exhibits a much broader OAM distribution than the bulk ppLN.

The corresponding marginal distributions of the OAM probabilities are shown in Figures.~\ref{fig:OAM_spectrum_both}(c,d). The experimental points (red) were obtained from the measured coincidence rates after correcting for the OAM-dependent detection efficiency. To enable a direct comparison with the theoretical prediction (blue bars), both distributions were normalized to the maximum.
Overall, the measured distributions are in good agreement with the theoretical ones. The remaining small deviations are likely caused by imperfections in the spatial-mode detection system, including mode-dependent diffraction and fiber-coupling efficiencies. In the case of the bulk sample, additional disagreement can be attributed to a small mismatch between the actual and nominal poling periods, which gives an additional phase mismatch and broadens the OAM spectrum \cite{Miatto2012Spatial}.

\subsection{Spatial--spectral decoupling}

\begin{figure}[h]
    \centering
    \includegraphics[width=1\linewidth]{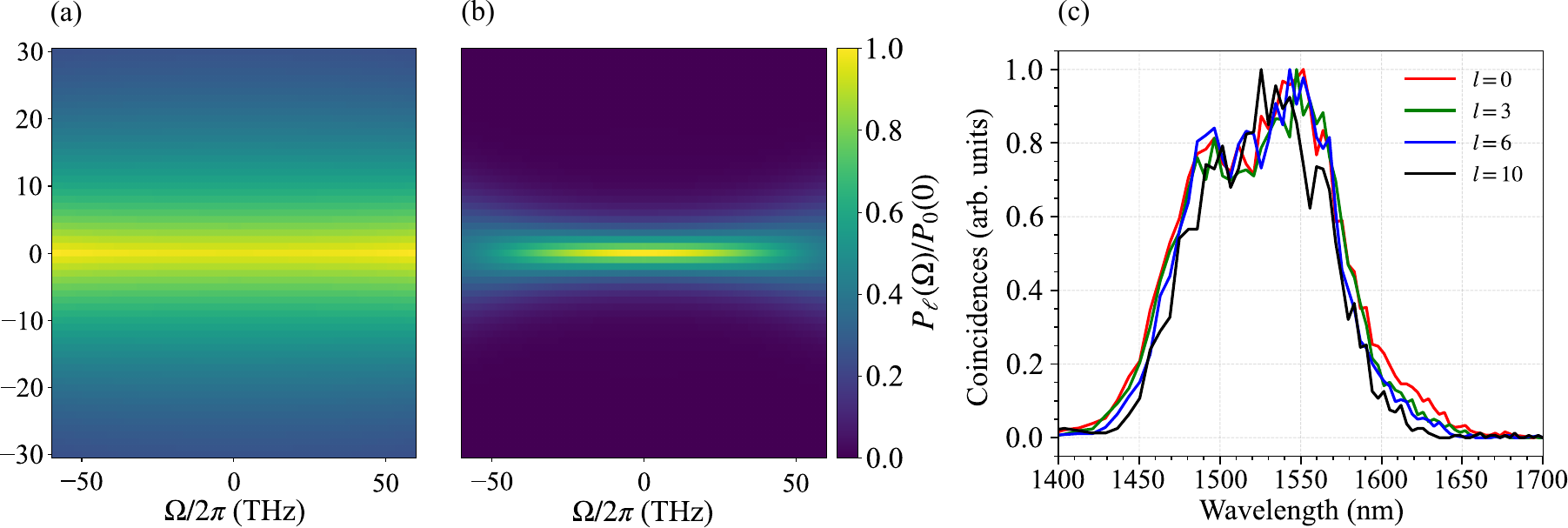}
    \caption{Calculated frequency-resolved OAM spectra of SPDC generated with a $775~\mathrm{nm}$ pump and a pump-beam waist of $w_p=37.5~\mu\mathrm{m}$ in (a) a $7~\mu\mathrm{m}$-thick LN film and (b) a $500~\mu\mathrm{m}$-thick ppLN crystal. Panel (c) shows the signal-photon frequency spectra for the LN film, measured for different OAM charges $\ell$.}
    \label{fig:freq_momentum_OAM}
\end{figure}

Figure~\ref{fig:freq_momentum_OAM} shows the frequency-resolved OAM spectra of the thin-film LN (a) and bulk ppLN crystal (b), calculated using the approach described in Section~3. All spectra were normalized to the probability of the $\ell=0$ mode at zero frequency detuning, $P_0(0)$. The thin LN film features a constant OAM spectrum across all the SPDC bandwidth, whereas the bulk ppLN exhibits a pronounced variation of the OAM distribution with frequency. This behaviour is consistent with spatial--spectral decoupling in the thin-film source, in contrast to the bulk PPLN crystal.

To experimentally investigate this behavior, we additionally performed frequency-resolved OAM measurements. A $3~\mathrm{km}$-long SMF-28 fiber was introduced into one detection arm as a dispersive element for time-of-flight spectroscopy. The results are presented in
Figure~\ref{fig:freq_momentum_OAM}(c). For OAM charges up to $|\ell|=10$, the measured spectral distribution remains nearly unchanged, providing experimental evidence of weak spatial--spectral
coupling over the investigated spectral range around the degenerate wavelength of $1550~\mathrm{nm}$. The experimentally accessible bandwidth is, however, limited by the detection scheme. On the short-wavelength side, time-of-flight spectroscopy becomes ineffective near the zero-dispersion wavelength of the SMF-28 fiber (around $1310~\mathrm{nm}$), where the wavelength-to-arrival-time mapping becomes weak. On the long-wavelength side, the measurement range is restricted by the operating range of the SLM (up to $1600~\mathrm{nm}$). Consequently, the overall detection bandwidth is relatively narrow compared with the broadband SPDC spectrum generated by the thin film.

Despite this technical limitation, the observed stability of the OAM spectrum across this range provides the first experimental indication of spatial--spectral decoupling in the thin-film source.

\subsection{Quantum tomography of OAM states}

The broader OAM spectrum of entangled photons from the nonlinear thin film provides access to a larger spatial Hilbert space. It therefore offers the potential for generating states of higher dimensionality than those produced by bulk crystals. However, the observation of a broad spiral bandwidth alone indicates only the population of multiple OAM modes but not their coherence.

To verify that the generated OAM components form coherent high-dimensional quantum states, we performed quantum-state tomography in selected $d$-dimensional subspaces~\cite{TomographyAgnew2011, Jack2009OAMTomography}. 
\begin{figure}
    \centering
    \includegraphics[width=0.7\linewidth]{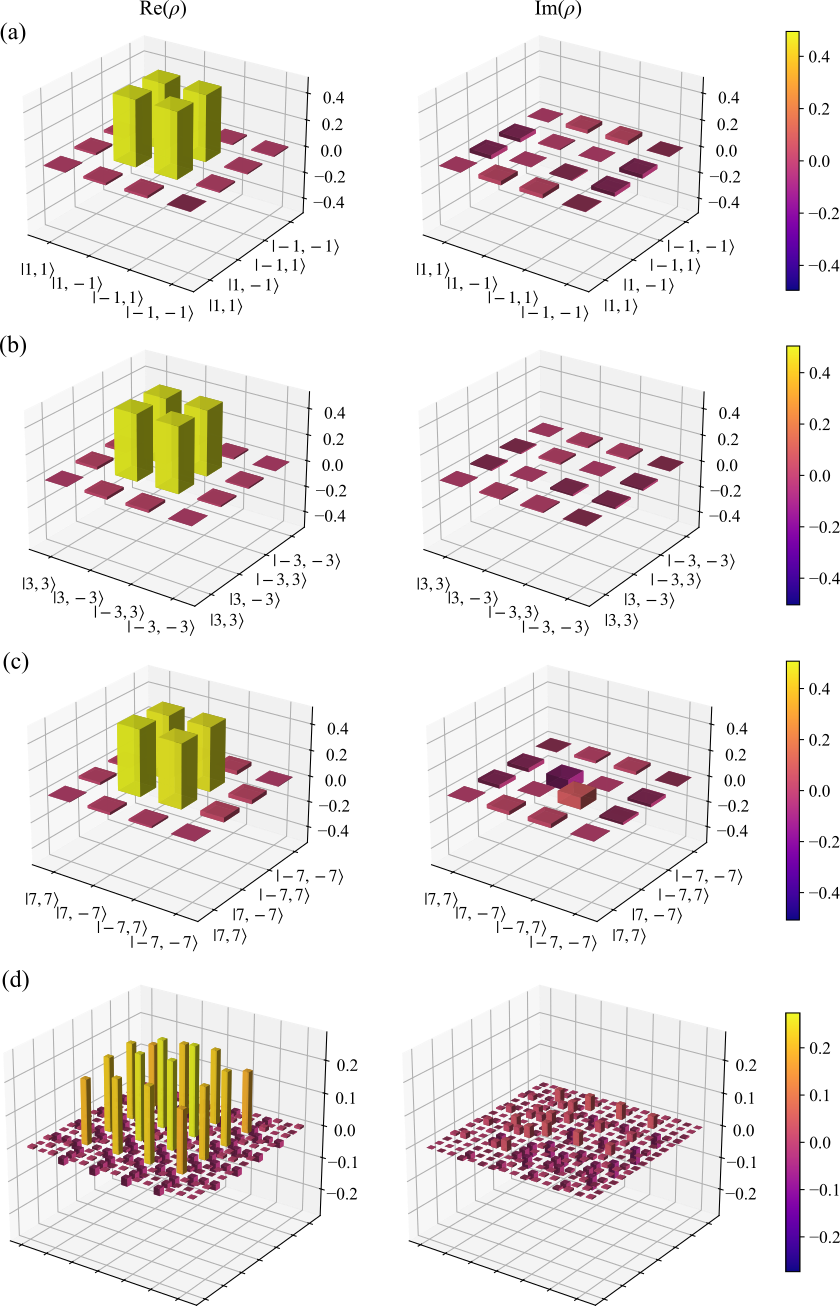}
    \caption{Real (left) and imaginary (right) parts of the reconstructed density matrices $\rho_d$ of the generated two-photon two-dimensional states (\ref{eq:2D})  with $\ell=1$ (a), $\ell=3$ (b), and $\ell=7$ (c), and four-dimensional state (\ref{eq:4D}) (d).
    }
    \label{fig:tomography_results}
\end{figure}
In particular, we reconstructed the density matrices $\rho_d$ of two-photon states in three two-dimensional OAM subspaces corresponding to the maximally entangled Bell states
\begin{equation}
    \ket{\Psi_{d=2}^{\ell}} = \frac{1}{\sqrt{2}} \left( \ket{\ell,-\ell} + \ket{-\ell,\ell} \right),
    \label{eq:2D}
\end{equation}
for $\ell=1,3, 7$. Here, $\ket{\ell_s,\ell_i} = \ket{\ell_s}_s\ket{\ell_i}_i$ denotes the two-photon OAM basis state.  
We also investigated the two-photon ququart state ($d=4$)

\begin{equation}
\ket{\Psi_{d=4}} = \frac{1}{2}\left( \ket{-2,2}+\ket{-1,1}+\ket{1,-1}+\ket{2,-2}\right),
\label{eq:4D}
\end{equation}
which represents a maximally entangled two-photon state in a four-dimensional OAM subspace. The reconstructed density matrices $\rho_d$ are shown in Figure~\ref{fig:tomography_results}.
From these density matrices, for each state we calculated the purity,
$
\mathrm{P} = \operatorname{Tr}\left(\rho_d^2\right),
$
and the fidelity to the corresponding target Bell state, summarized in Table~\ref{tab:state-purity-fidelity}.
\begin{table}[htbp]
\centering
\caption{Purity and fidelity of the generated two-photon states $\ket{\Psi_{d=2}^{\ell}}=\frac{1}{\sqrt{2}}\left(\ket{\ell,-\ell}+\ket{-\ell,\ell}\right)$ and $\ket{\Psi_{d=4}}$, corresponding to a four-dimensional Bell-like state.}
\label{tab:state-purity-fidelity}

\begin{tabular}{ccccc}
\hline
& $\ket{\Psi_{d=2}^{1}}$ & $\ket{\Psi_{d=2}^{3}}$ & $\ket{\Psi_{d=2}^{7}}$ & $\ket{\Psi_{d=4}}$ \\
\hline
$\mathrm{F}$ & $0.981\pm0.009$ & $0.996\pm0.002$ & $0.982\pm0.008$ & $0.916\pm0.014$\\
$\mathrm{P}$  & $0.986\pm0.005$ & $0.997\pm0.002$ & $0.99\pm0.01$ & $0.956\pm 0.025$ \\
\hline
\end{tabular}
\end{table}

For the three two-dimensional OAM subspaces, the reconstructed states show excellent agreement with the corresponding Bell states, with fidelities exceeding $0.98$ and purities close to unity. Notably, this high state quality is maintained for widely separated OAM orders, indicating that high-quality two-photon superpositions can be extracted from different regions of the broad spiral spectrum.

For the four-dimensional subspace, the reconstructed state retains purity of $\mathrm{P}=0.956$ and a fidelity of $\mathrm{F}=0.916$ with respect to the target maximally entangled
Bell-like state. The lower fidelity compared to the two-dimensional states can be attributed to the increased complexity of the higher-dimensional reconstruction, which involves a larger number of mode projections and is more sensitive to mode-dependent diffraction and coupling efficiencies, residual cross-talk, and counting-statistics fluctuations.

Importantly, the simultaneous observation of a broad OAM distribution and high-fidelity Bell-type states shows that the large accessible spatial dimensionality of the thin-film source is accompanied by high quantum-state quality.

\section{Conclusion}

In conclusion, we have shown that thin-film SPDC supports not only broadband spectral and angular emission due to the relaxed phase-matching conditions, but also a substantially broadened OAM spectrum. Using a $7~\mu\mathrm{m}$-thick lithium niobate film, we have demonstrated a high spatial dimensionality of the generated photon pairs, reaching an azimuthal Schmidt number of $K_{\mathrm{az}}^{\mathrm{LN}}=59.4$, almost nine times higher than for a $500~\mu\mathrm{m}$-thick phase-matched ppLN crystal under identical conditions.
Within this high-dimensional OAM space, we have performed quantum-state tomography in selected two-dimensional subspaces corresponding to Bell states with $\ell=1,3,$ and $7$, as well as in a four-dimensional OAM subspace. The reconstructed states exhibit high purity and high fidelity to the corresponding maximally entangled target states.

We further investigated the spatial--spectral properties of the generated OAM states. According to numerical simulations, predicting an essentially unchanged OAM spectrum over a broad spectral range, our frequency-resolved OAM measurements showed no dependence of the OAM distribution on frequency within the accessible detection bandwidth. In other words, unlike a bulk SPDC source, a thin film effectively decouples the spatial and spectral entanglement within a very broad spectral range.
\section*{Acknowledgment}
\section*{Funding}
European Research Council (Project 101199215 — MultiFlaQS); Bundesministerium für Bildung und Forschung (13N16933); QM3 (Quantum Multi-Modal Microscopy) project under the QuantERA II programme (grant no. QuantERAII/02/QM3/03/2024), which received funding from the European Union’s Horizon 2020 Research and Innovation Programme under Grant Agreement No. 101017733; Foundation for Polish Science, the FIRST TEAM FENG programme (Project No. FENG.02.02-IP.05-0253/23).
\section*{Data availability}
Data underlying the results presented in this paper are not publicly available, but may be obtained from the authors upon reasonable request.

\bibliography{sample}






\end{document}